\documentclass[
amsmath,
amssymb,
showpacs,
twocolumn,
aps,prb]{revtex4-1}
\usepackage[]{graphicx}			
\usepackage{dcolumn}			
\usepackage[colorlinks=true, pdfstartview=FitV, linkcolor=blue, citecolor=blue, urlcolor=blue]{hyperref}
\usepackage{hyperref}
\usepackage{SIunits}

\begin{document}

\title{Magnetization reversal of an individual exchange biased permalloy nanotube}

\author{A. Buchter$^1$, R. W\"{o}lbing$^2$, M. Wyss$^1$, O. F. Kieler$^3$, T. Weimann$^3$, J. Kohlmann$^3$, A. B. Zorin$^3$, D. R\"{u}ffer$^4$, F. Matteini$^4$, G. T\"{u}t\"{u}nc\"{u}oglu$^4$, F. Heimbach$^5$, A. Kleibert$^6$, A. Fontcuberta i Morral$^4$, D. Grundler$^{5,7}$, R. Kleiner$^2$, D. Koelle$^2$, M. Poggio$^1$}

\affiliation{ $^1$ Department of Physics, University of Basel, 4056
  Basel, Switzerland; $^2$ Physikalisches Institut and Center for
  Quantum Science (CQ) in LISA$^+$, Universit\"{a}t T\"{u}bingen,
  72706 T\"{u}bingen, Germany; $^3$ Fachbereich „Quantenelektronik“,
  Physikalisch-Technische Bundesanstalt, 38116 Braunschweig, Germany;
  $^4$ Laboratoire des mat\'{e}riaux semiconducteurs LMSC de
  l'institut des mat\'{e}riaux IMX et Section de science et g\'{e}nie
  des mat\'{e}riaux SMX, Facult\'{e} des sciences et techniques de
  l'ing\'{e}nieur STI, Ecole polytechnique f\'{e}d\'{e}rale de
  Lausanne EPFL, 1015 Lausanne, Suisse; $^5$ Lehrstuhl f\"{u}r Physik
  funktionaler Schichtsysteme, Physik Department E10, Technische
  Universit\"{a}t M\"{u}nchen, 85747 Garching, Deutschland; $^6$ Swiss
  Light Source, Paul Scherrer Institute, 5232 Villigen PSI,
  Switzerland.  $^7$ Laboratory of Nanoscale Magnetic Materials and
  Magnonics LMGN, Institute of Materials IMX, Facult\'{e} des sciences
  et techniques de l'ing\'{e}nieur STI, Ecole polytechnique
  f\'{e}d\'{e}rale de Lausanne EPFL, 1015 Lausanne, Suisse; }

\date{\today}

\begin{abstract}
  We investigate the magnetization reversal mechanism in an individual
  permalloy (Py) nanotube (NT) using a hybrid magnetometer consisting
  of a nanometer-scale SQUID (nanoSQUID) and a cantilever torque
  sensor.  The Py NT is affixed to the tip of a Si cantilever and
  positioned in order to optimally couple its stray flux into a Nb
  nanoSQUID.  We are thus able to measure both the NT's volume
  magnetization by dynamic cantilever magnetometry and its stray flux
  using the nanoSQUID.  We observe a training effect and temperature
  dependence in the magnetic hysteresis, suggesting an exchange bias.
  We find a low blocking temperature $T_B=18\pm 2$ K, indicating the
  presence of a thin antiferromagnetic native oxide, as confirmed by
  X-ray absorption spectroscopy on similar samples.  Furthermore, we
  measure changes in the shape of the magnetic hysteresis as a
  function of temperature and increased training.  These observations
  show that the presence of a thin exchange-coupled native oxide
  modifies the magnetization reversal process at low temperatures.
  Complementary information obtained via cantilever and nanoSQUID
  magnetometry allows us to conclude that, in the absence of exchange
  coupling, this reversal process is nucleated at the NT's ends and
  propagates along its length as predicted by theory.

\end{abstract}


\maketitle

Fabrication and characterization of magnetic nanostructures is
motivated by a wide range of applications, including their use as
media in dense magnetic memories \citep{Parkin:2008}, as magnetic
sensors \citep{Maqableh:2012}, or as probes in high resolution
magnetic imaging \citep{Campanella:2011, Poggio:2010,
  Khizroev:2002}. The desire for higher density memories, more
sensitive sensors, and higher resolution imaging has pushed magnet
size deep into the nanometer-scale.  At these length scales, the
stability of magnetization configurations strongly depends on
geometry, defects, and minute levels of contamination.  This
sensitivity to imperfection makes the experimental realization of
idealized systems such as ferromagnetic rods and tubes particularly
challenging.  Furthermore, due to the small total magnetic moment of
each nanomagnet, conventional magnetometry techniques do not have the
necessary sensitivity to measure individual nanostructures.  As a
result, measurements of their magnetic properties are often carried
out on large ensembles, whose constituent nanomagnets have a
distribution of size, shape, and orientation and -- depending on the
density -- may interact with each other \citep{Escrig:2008,
  EscrigAPL:2008}.  These complications conspire to make accurate
characterization of the stable magnetization configurations and
reversal processes difficult.

In order to obtain a clear understanding of the magnetic properties of
ferromagnetic nanotubes (NTs), it is therefore advantageous to
investigate individual specimens.  Ferromagnetic NTs are particularly
interesting nanomagnets because of their lack of a magnetic core.
This geometry can make flux-closure magetization configurations more
favorable than single-domain states \citep{Escrig:2007}.  Flux-closure
configurations are predicted to enable fast and reproducible
magnetization reversal and they produce minimal stray magnetic fields,
thereby reducing interactions between nearby nanomagnets.  We
therefore measure the magnetization and stray field hysteresis of an
individual permalloy (Py) NT using a hybrid magnetometer.  The
magnetometer combines a sensitive mechanical sensor for dynamic
cantilever magnetometry (DCM) and a nanometer-scale SQUID (nanoSQUID)
for the measurement of stray magnetic fields.  This measurement
technique was first demonstrated on individual Ni NTs by Buchter et
al.\citep{Buchter:2013}, who revealed the importance of morphological
defects in altering the reversal process in real ferromagnetic NTs
from the theoretical ideal.

\begin{figure*}[t]
  \centering
  \includegraphics[width=2\columnwidth]{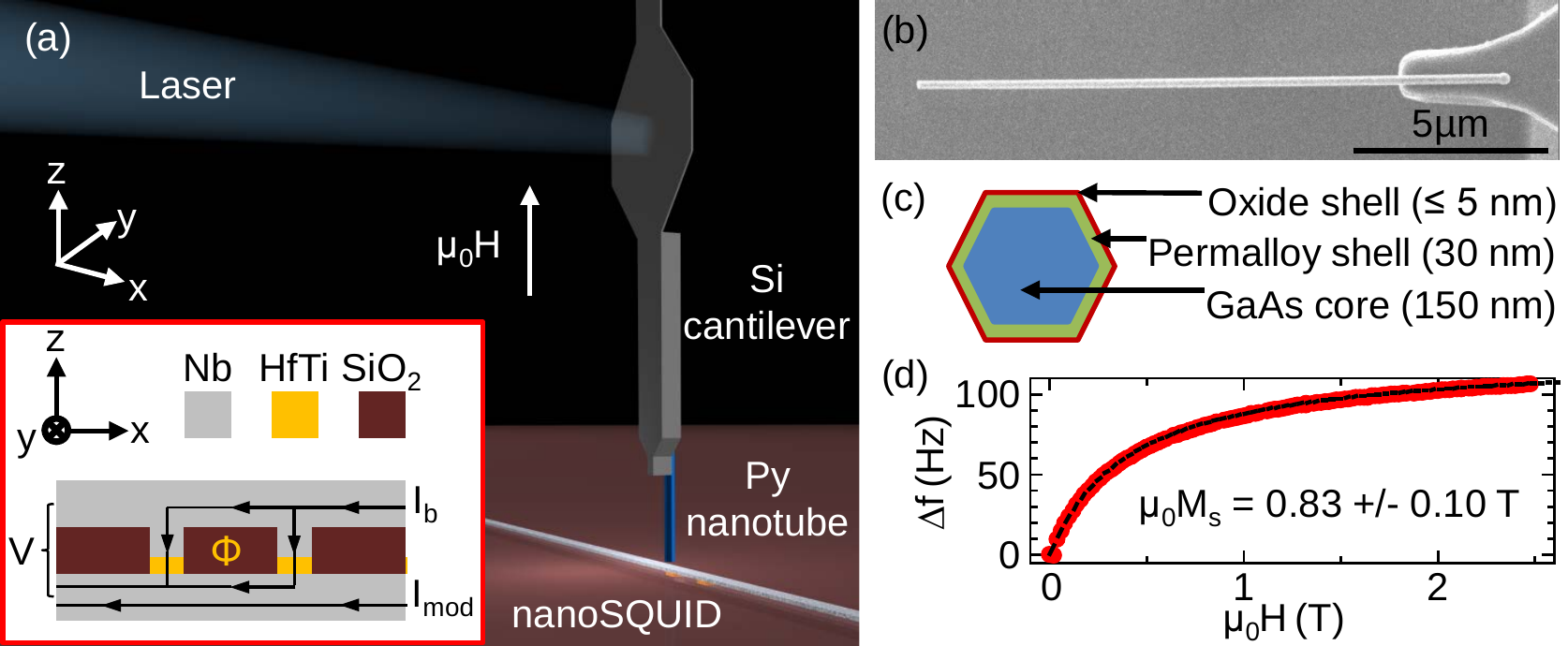}
\caption{
\label{fig1}
(a) Schematic of the hybrid magnetometer.  Inset: cross-sectional
diagram of the nanoSQUID showing the voltage $V$, bias current $I_b$,
and modulation current $I_{mod}$. (b) SEM of the investigated Py NT on
cantilever.  (c) Schematic cross-section of a Py NT as inferred from
the measurements. (d) Measurement of $\Delta f (H)$ of a Py NT by
DCM. The saturation magnetization $\mu_0 M_s = 0.83 \pm 0.10$ T is
determined by a fit (dashed line) to the data (filled circles)
(\ref{eq:lowFieldLimit}). }
\end{figure*}

Here, we study individual Py NTs. The fabrication process is based on
evaporation instead of atomic layer deposition as used for Ni NTs
\citep{Ruffer:2012} and provides polycrystalline Py NTs with smooth
surface that are morphologically closer to an idealized tube. Despite
the geometrical perfection of the Py NTs, the measured low-temperature
hysteresis curves reveal, that a thin exchange-coupled native oxide
changes the reversal process \citep{Meiklejohn:1956}.  Since the oxide
is thin -- likely less than 5 nm -- these effects only appear at
temperatures below 20 K.  The role of the oxide is only apparent due
to the sensitivity of the hybrid magnetometer to single NTs, since
averaging effects would likely obscure the behavior in conventional
measurements of NT ensembles.  The strong effect of such a thin oxide
layer on magnetic reversal, points to the importance of gaining
further control of the fabrication of ferromagnetic NTs.  At the same
time, the results indicate that engineered oxide layers could be used
to pin or otherwise control the magnetic configurations of magnetic
NTs.

In order to fabricate the Py NTs, GaAs nanowires grown by molecular
beam epitaxy are used as templates.  These nanowires
are 10 to 20 $\mu$m long and have hexagonal cross-sections with a
width $d=150 \pm 20$ nm at their widest point.  Tubes of magnetic
material are formed by thermally evaporating a $30$ nm polycristalline
Py shell onto the template nanowires.  For this deposition, the low
density GaAs nanowire wafer is mounted under 35$^\circ$ angle and
continuously rotated in order to achieve a conformal coating. The films fabricated in this process are very smooth, show no discontinuities and the roughness is less then $5$ nm.  Individual Py NTs are then
selected and transferred from the wafer to the end of a Si cantilever
using high precision hydraulic micro-manipulators (Narishige
MMO-202ND) mounted under an optical microscope.  Each NT is attached
parallel to the cantilever's long axis, such that it protrudes from
its end by $\sim 12$ $\mu$m.  The NT studied in this work is $14.8$
$\mu$m long, and has a magnetic volume $V=(2.46 \pm 0.18) \cdot 10^{-19}$
m$^3$.

In our hybrid magnetometer, the single-crystal Si cantilever serves
two purposes: 1) as the torque transducer in DCM measurements of the
NT and 2) as a platform for positioning the NT such that stray
magnetic fields from its tip couple to the nanoSQUID.  It is $105$
$\mu$m long, $4$ $\mu$m wide and $0.1$ $\mu$m thick and has a $1$
$\mu$m thick and $18$ $\mu$m long mass at its end to suppress higher
order modes \citep{Chui:2003}.  The NT-tipped cantilever hangs in the
pendulum geometry above the nanoSQUID in a vacuum chamber at the
bottom of a temperature variable He$^3$ cryostat.  This system is
capable of temperatures down to $0.3$ K and magnetic fields up to $6$
T along the long axis of the cantilever ($\hat{z}$).  The read-out of
the cantilever deflection is achieved through a laser interferometer.
We use a temperature controlled $1550$ nm fiber-coupled DFB laser
diode and an interferometer cavity formed between the cleaved end of
an optical fibre and the $12$ $\mu$m wide paddle of the Si cantilever
\citep{Rugar:1989}.  By feeding the measured displacement signal
through a field-programmable gate array (National Instruments) and
back to a piezo-electric actuator mechanically coupled to the
cantilever, we self-oscillate the cantilever at its mechanical
resonance frequency $f_c$.  The oscillation amplitude is stabilized to
$x_{rms} = 10$ nm, for which the deflection angle $\theta_{rms} \ll
0.1^{\circ}$, allowing for fast and accurate determination of $f_c$.
At a temperature of $3.8$ K and far from any surface the cantilever
has an intrinsic resonance frequency of $f_c=f_0=3980$ Hz and a
quality factor $Q_0=41\cdot 10^3$.  The spring constant $k_0=185$
$\mu$N/m is determined by thermal noise measurements at different
temperatures.

In order to control the relative position of the NT and the nanoSQUID,
the nanoSQUID is mounted on a three dimensional piezoelectric
positioning stage (Attocube).  We use a direct current nanoSQUID
containing two superconductor-normal metal-superconductor Josephson
junctions (JJs) in a microstrip geometry \citep{WoelbingAPL:2013}.
Two $250$ nm wide and $200$ nm thick Nb strips lie on top of each
other and are separated by a $224$ nm thick SiO$_2$ insulating layer.
To form the $1.6 \times 0.224$ $\mu$m$^2$ SQUID loop, the Nb strips
are connected by two Nb/HfTi/Nb JJs with $200 \times 200$ nm$^2$ area
and a $24$ nm thick HfTi barrier.  Using a cryogenic series SQUID
array as a low noise amplifier in a magnetically and electrically
shielded environment and Magnicon XXF-1 read-out electronics, the
described nanoSQUID shows a white rms flux noise $S_\Phi^{1/2}=190$
n$\Phi_0$/Hz$^{1/2}$ between 1 and 10 kHz.  Here,
$\Phi_0=\frac{h}{2e}\approx 2.07 \times 10^{-15}$ $\text{V} \cdot
\text{s}$ is the magnetic flux quantum.  At lower frequencies, the
noise has a 1/f-like spectrum with a corner frequency around 200 Hz
(see supplementary material).

We first investigate the NT sample through high-field DCM as plotted
in Fig.~\ref{fig1}d).  The field $\mu_0 H$ is swept from $0$ T to
$2.5$ T in $20$ mT increments, while the frequency shift of the
cantilever $\Delta f(H)=f_c(H)-f_0$ is measured. The NT is composed of
a Py shell that is known to be magnetically isotropic at room
temperature. At small $T$, the polycrystalline morphology is expected
to average out any magnetocrystalline anisotropy.  Given the latter
and given the large aspect ratio of the NT, its anisotropy energy is
dominated by shape effect.  Therefore, the data are fit to an
analytical model describing a Stoner-Wohlfarth particle with shape
anisotropy \citep{Weber:2012}. In this model, the NT is idealized as a
uniformly magnetized magnet whose magnetization rotates in unison.
For $H$ applied along the easy axis of a sample with large shape
anisotropy -- as in this case -- the magnetization is forced to be
either parallel or anti-parallel to the applied field.  As a result,
this model is an excellent approximation for most of the field range,
except in the regions of magnetic reversal.  The model predicts a DCM
frequency shift given by,
\begin{equation}
  \Delta f = \frac{f_0 \mu_0 V}{2 k_0 l_e^2} \left ( \frac{-D_u M_s^2 H}{H \mp D_u M_s} \right ),  
\label{eq:lowFieldLimit}
\end{equation}
where $\mu_0$ is the permeability of free space, $V$ is the volume of
the Py NT, $l_e$ is the effective length of the cantilever for its
fundamental mode, $M_s$ is the saturation magnetization, and $D_u$ is
the effective uniaxial demagnetization factor along $\hat{z}$
\citep{Weber:2012,Gross:2015}.  The two solutions are valid for $H >
D_u M_s$ and $H < -D_u M_s$ respectively, which for the NT's easy-axis
anisotropy ($D_u < 0$) results in a region of bistability, allowing
for magnetic hysteresis.  By fitting the measurements shown in
Fig.\ref{fig1}d) with this expression, we extract $\mu_0 M_s=0.83 \pm
0.10$ T and $D_u= -0.496 \pm 0.001$ as fit parameters.  Input
parameters are $f_0$, $k_0$, $l_e$, and $V$, which are all set to
their measured values.  The saturation magnetization measured for the
Py NT is smaller than the literature value $\mu_0 M_{Py}=1$ T
\citep{Thiaville:2005} for bulk Py. This discrepancy may be the result
of an overestimation of the NT volume due to the oxide layer present
on the surface or due to other imperfections in the growth of the
film. The measured demagnetization factor, however, is in excellent
agreement with what can be calculated by approximating the NT as a
hollow cylinder, ignoring its hexagonal cross-section.  The main
contribution to the error in the extracted values stems from the NT
volume, which is difficult to determine precisely.  We calculate $V$
using the NT dimensions extracted from scanning electron micrographs
(SEMs) of the Py shell thickness. A further source of error is the
determination of the cantilever's spring constant.

We now turn to the low-field behavior of the sample and the
investigation of exchange bias.  As key feature of exchange bias
systems, we start by exploring the training effect of our sample.  For
initialization, the sample chamber is heated above $110$ K.  The
subsequent cool-down to $3.4$ K is done with an applied magnetic field
of $+200$ mT, in order to create a defined state of magnetization in
the NT.  Exploiting the duality of our hybrid magnetometer, we measure
both the stray magnetic flux generated by the NT with the nanoSQUID
and the integrated magnetization by DCM.  For the nanoSQUID
measurements, the NT-tipped cantilever is positioned for optimal
coupling at height $z=1.1$ $\mu$m above the nanoSQUID's top electrode
\citep{NagelPRB:2013,Buchter:2013}.  Despite this proximity of the
nanoSQUID to the NT, the magnetic fields produced by the bias and
modulation currents running through the nanoSQUID and its
superconducting leads are much less than $1$ mT at the position of the
NT tip and do not significantly influence its magnetization state.
Measurements of DCM, on the other hand, are carried out with the NT
several tens of $\mu$m away from the nanoSQUID in $\hat{z}$. This
large spacing avoids spurious magnetic torque generated by the
magnetic field gradients of the nanoSQUID, which otherwise complicate
conversion of $\Delta f(H)$ measured by DCM into magnetization $M(H)$.

In both cases, the magnetic field is swept first from $0$ mT to $+55$
mT and subsequently between $ \pm 55$ mT until ten full hysteresis
loops are completed.
\begin{figure}[t]
  \centering
  \includegraphics[width=1\columnwidth]{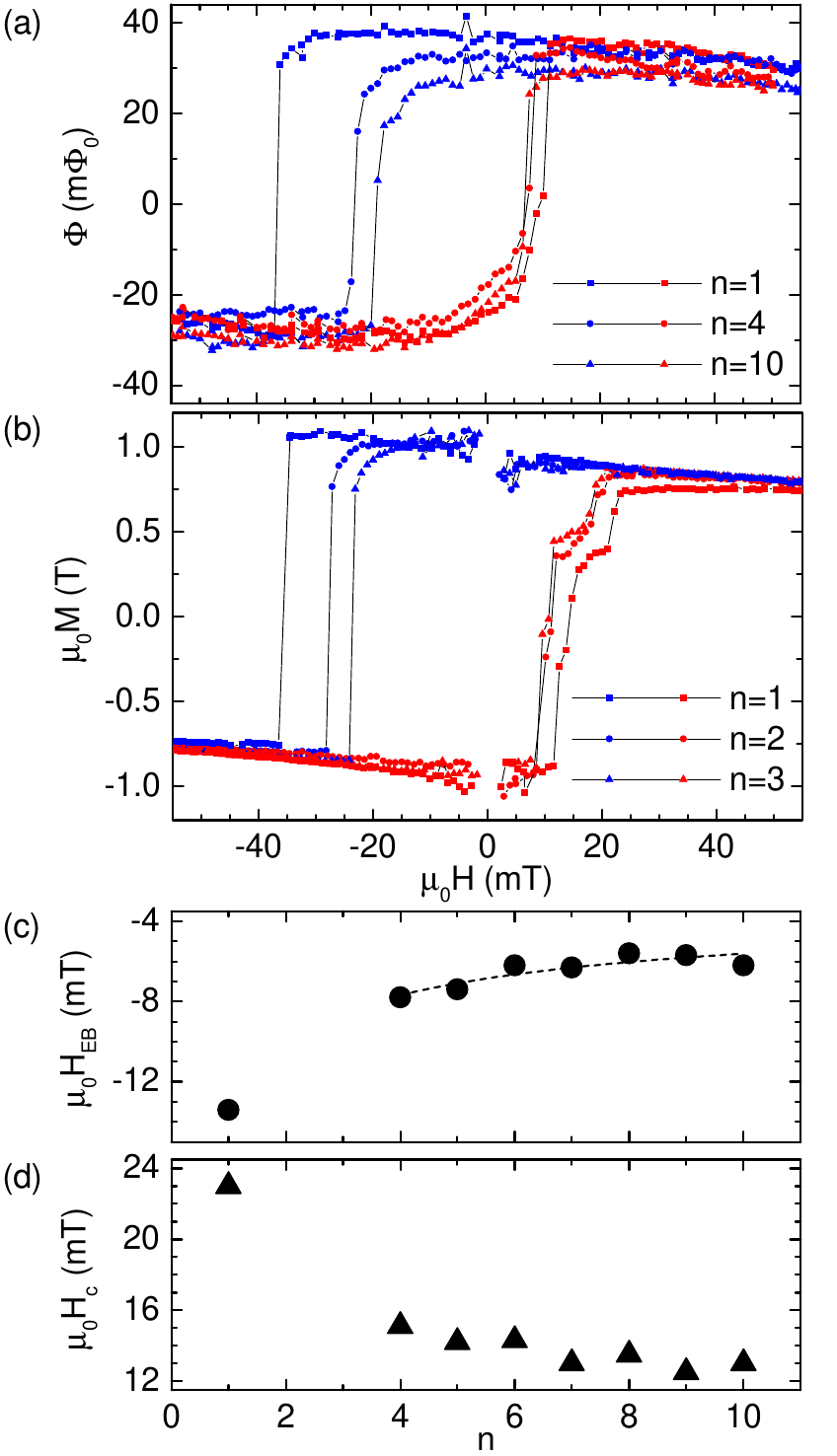}
\caption{
\label{fig2}
Training effect: (a) SQUID and (b) DCM hysteresis loops for different
loop number $n$ at $T = 3.4$ K. Red and blue curves indicate up- and
down-sweep respectively. Evolution of (c) exchange field and (d)
coercivity with increasing loop number $n$ extracted from SQUID
dataset.  Dashed line fits the the data according to
Eq. (\ref{eq:training_sqrt}). Point size corresponds to the
measurement error in field. }
\end{figure}
Fig.~\ref{fig2}a) shows several iterations of magnetic hysteresis of
the stray flux emanating from the NT measured by the nanoSQUID. In
order to isolate the magnetic flux emerging from the Py NT from the
spurious flux due to the external field threading through the slightly
misaligned nanoSQUID, we record a reference sweep with the NT
retracted tens of $\mu$m from the substrate.  These reference data,
which exclude effects due to the NT, are then used both to subtract
spurious flux and as a calibration of the magnetic field-axis
\citep{Buchter:2013}.  The blue squares show the first sweep from $55$
mT to $-55$ mT.  The flux from the NT starts at $30$ m$\Phi_0$ and
slightly increases to $38$ m$\Phi_0$ at $-25$ mT.  This measurement
artefact is the result of a slight temperature difference between the
$n=1$ loop and the reference sweep.  Thermal drift developing in the
timespan of approximately $10$ hours between the two measurements
alters the nanoSQUID's response.  Magnetization reversal takes place
in a field range from $-31.1$ mT to $-36.9$ mT with a coercive field
of $\mu_0 H_{c-}=-36.5$ mT.  The reversal sets in by a steady decrease
in flux of $6.9$ m$\Phi_0$ followed by a large, abrupt change of
$59.6$ m$\Phi_0$.  Following the subsequent up-sweep (red squares),
the temperature-induced slope can again be identified.  In this case,
reversal takes place over a wider field range than in the down-sweep,
from $-12.0$ mT to $11.2$ mT with a coercive field $\mu_0 H_{c+}=9.4$
mT.  We first observe a steady decrease of $9.0$ m$\Phi_0$ of flux
followed by a region of steeper change in flux (in the range of $6$
m$\Phi_0$), and ending with a final discontinuity of $33$ m$\Phi_0$.
The total width of the loop is identified as the coercivity $\mu_0
H_c=\mu_0 (H_{c+}-H_{c-})/2=23$ mT.  We attribute the strong asymmetry
of the hysteresis along the field axis to the exchange bias effect as
will be detailed below.  The corresponding parameter, the
exchange-bias field, is defined as $\mu_0
H_{EB}=\mu_0(H_{c+}+H_{c-})/2=-13.5$ mT.

Subsequent sweeps show similar behavior, although with a progressive
decrease in $|\mu_0 H_{c-}|$, as seen in the representative sweeps of
Fig.~\ref{fig2}a).  Following the blue triangles of the $n=10$ sweep,
we observe a smaller thermal drift effect than in the $n=1$-loop due
to the proximity in time between the $n=10$-loop and the reference
sweep taken at the end of the series.  On the down-sweep, from $0$ mT
to $-17.8$ mT, the flux coupled into the nanoSQUID steadily decreases
by $12.5$ m$\Phi_0$, followed by two abrupt switching events until the
reversal process completes at $-20.0$ mT.  The reversal process on the
up-sweep starts at $-14.2$ mT with a continuous reduction of stray
field of $14.7$ m$\Phi_0$ over a range of $9.0$ mT.  This continuous
reversal is then followed by abrupt switching events until the process
completes at $10$ mT.  The most striking deviations from the first
loop are the reduced width of the hysteresis loop -- now $\mu_0
H_c=13.0$ mT -- and the reduction of the exchange bias field $\mu_0
H_{EB}=-6.2$ mT. These findings are summarized in Fig.~\ref{fig2}c),
where $\mu_0 H_{EB}$ and $\mu_0 H_c$ for $n=1$ to $10$ are
plotted. $\left| \mu_0 H_{EB} \right| $ decreases by $\sim 7.2$ mT
until $n=6$, after which point it stabilizes at $-6\pm 0.5$ mT.
$\mu_0 H_c$ reduces by $\sim 10$ mT within the first seven loops until
a saturation at $13 \pm 0.5$ mT.  Past studies showed that the
evolution of $\mu_0 H_{EB}$ can be described by the following formula
especially in the case of a polycrystalline antiferromagnet
\citep{NoguesEB:1999, Binek:2004, Paccard:1966}:
\begin{equation}
  \mu_0 H_{EB}^n - \mu_0 H_{EB}^{\infty} = \frac{\kappa}{\sqrt{n}},  
\label{eq:training_sqrt}
\end{equation}
where $\kappa$ is a system dependent parameter and $\mu_0 H_{EB}^n$
and $\mu_0 H_{EB}^{\infty}$ are the exchange bias fields after $n$
loops and in the limit of infinite number of loops respectively. The
fit for $3<n<11$ is shown as dashed line in Fig. \ref{fig2}c). The
data are reasonably described by this power-law with deviations likely
due to the inhomogeneity (in composition and thickness) of the
naturally oxidized antiferromagnetic layer. As fit parameters we
obtain $\kappa= -11.3$ mT and $\mu_0 H_{EB}^{\infty}= -2.0$ mT, whose
magnitudes are of the same order as seen in literature
\citep{ProencaPRB:2013}. Note also that as a function of training
(increasing $n$), the hysteresis loop becomes more symmetric, losing
the difference in the shape of the magnetization reversal for up- and
down-sweeps.  In particular, the abrupt reversal seen on the initial
down-sweep is in stark contrast with the rounded transitions of later
sweeps.

After an identical initialization and field-cooling procedure, DCM
data are taken further away from the nanoSQUID, under otherwise
identical measurement conditions.  The magnetization hysteresis, shown
in Fig.~\ref{fig2}b), can be extracted from measurements of $\Delta
f(H)$ at low applied magnetic fields, by taking the limit of
(\ref{eq:lowFieldLimit}) for $H \ll D_u M_s$ and solving for $M_z$:
$M_z = \frac{2 k_0 l_e^2}{f_0 \mu_0 V H} \Delta f$ \citep{Gross:2015}.
Such data reflect the integrated magnetization of the entire NT,
rather than the magnetization of the NT end closest to the substrate,
as do the nanoSQUID measurements.  The behavior of $H_{EB}$ and $H_c$
reproduces what we observe with the nanoSQUID.  Nevertheless,
differences are observable in the appearance of the magnetization
reversals.  For the $n=1$ down-sweep, the single step magnetization
reversal measured by DCM resembles that measured by the nanoSQUID.
The reversal on the up-sweep shows a two-stage behavior, including an
intermediate plateau, different than the continuous reversal followed
by an abrupt switching seen in the corresponding nanoSQUID data.  For
$n>1$, the down-sweep reversal gradually develops an initial stage of
coherent reversal before discontinuous switching.  Most strikingly,
the coherent reversal seen in the nanoSQUID measurements, precedes the
beginnings of any reversal observed by DCM.  The discontinuous steps
seen in the nanoSQUID sweeps coincide with the first discontinuous
steps observed in DCM.  The plateau and second discontinous reversal
measured in DCM corresponds to a portion of the nanoSQUID hysteresis
that has already reached saturation.

These findings lead us to two conclusions. First, the differences
observed in the hysteresis measured by the nanoSQUID and by DCM
indicate that magnetization reversal likely nucleates at the ends of
the NT and subsequently propagates throughout its length, as predicted
by theory \citep{Landeros:2009}. In this picture, vortex
configurations form at the NT ends, begin tilting in the direction of
the applied field, and subsequently cause magnetization reversal by
propagating throughout the length of the tube and discontinuously
switching to a uniform configuration aligned along the applied field.
This kind of reversal is consistent with nanoSQUID measurements that
show a smooth reversible reduction in stray field, which precede any
deviation of $M_z$ from saturation registered by DCM.  The subsequent
irreversible change in the hysteresis is then registered both in the
nanoSQUID and DCM measurements, thus apparently occuring throughout
the NT and not only at the ends.  The final stages of reversal seen in
DCM, i.e.\ the plateau and second discontinuous step, appear to occur
far from the NT end, since at this stage the nanoSQUID already shows a
saturated signal.  In short, the measurements appear consistent with
the theoretical picture of reversal nucleation at the NT ends,
considering that the nanoSQUID is sensitive to magnetization located
at the NT end and DCM is sensitive to the total magnetization
integrated throughout the NT.  Unlike the Ni NTs measured by Buchter
et al. \citep{Buchter:2013}, whose reversal did not nucleate at the
ends most likely due to imperfections in their structure, these Py NTs
appear to behave like the idealized magnetic NTs considered in
theoretical calculations.

Second, the exchange bias effect influences the nature of the reversal
process of the NT, as manifested in the changing shape of the
hysteresis as a function of training.  In particular, for small $n$,
the down-sweep magnetization reversal occurs almost exclusively
through a single irreversible change, while the up-sweep reversal and
both reversals for large $n$ contain both reversible rotation of
magnetization and irreversible switching.  The antiferromagnetic
layer, in its initial configuration, appears therefore to pin the
magnetization of the NT on the down-sweep, favoring reversal by abrupt
domain nucleation and propagation.  The exchange-coupled layer may
thus suppress nucleation and initial coherent reversal through vortex
cofigurations at the NT's ends.

\begin{figure}[b]
  \centering
  \includegraphics[width=1\columnwidth]{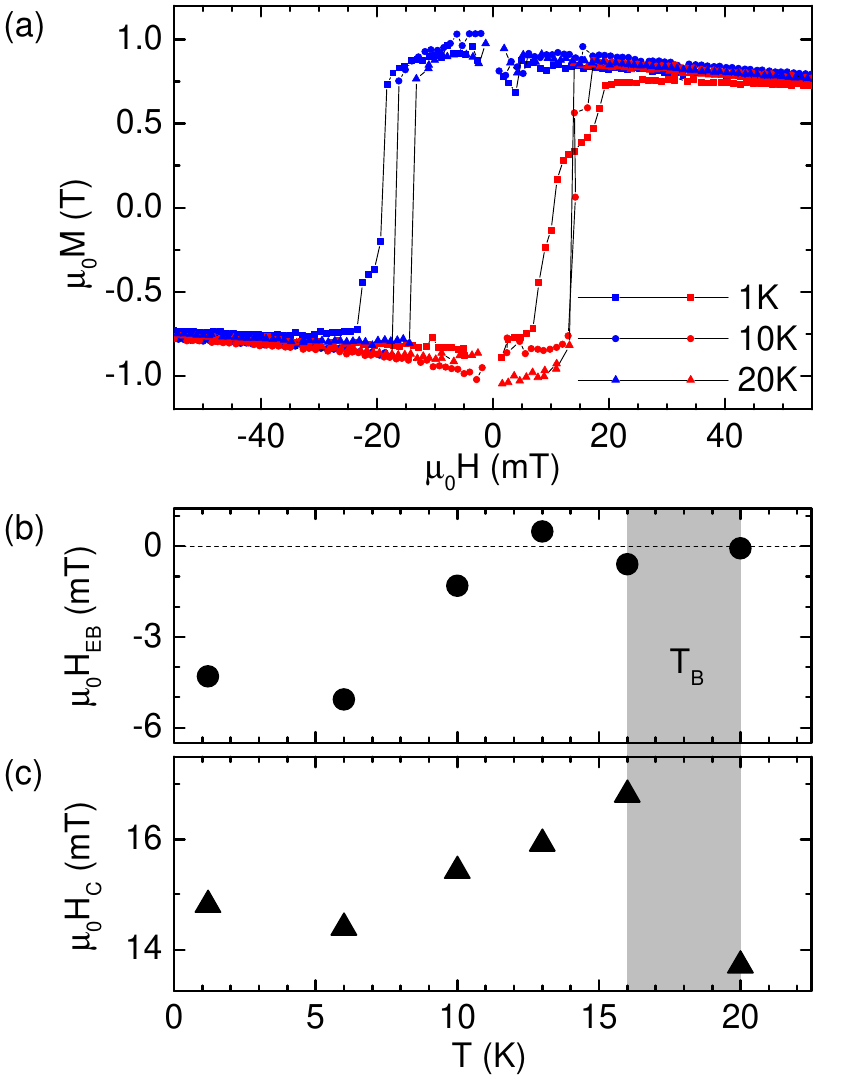}
\caption{
\label{fig3}
Temperature dependence: (a) DCM hysteresis loops for different
temperatures in a trained out state ($n>10$).  Evolution of (b)
exchange field and (c) coercivity with temperature.  Gray area
indicates range of $T_B$.  Point size corresponds to the measurement
error in field. }
\end{figure}

We next investigate the temperature dependence of the exchange bias
effect in the Py NT.  Hysteresis loops are measured by DCM in a
temperature range between $1$ K and $20$ K.  Concurrent measurements
with the nanoSQUID are not possible since the device's performance is
strongly temperature dependent and above the critical temperature of
Nb ($T_C \approx 9$ K), operation of the nanoSQUID is impossible.  In
order to ensure that our measurements are not obscured by training
effects, we measure the temperature dependence after the extensive
training of the NT, such that $H_{EB}$ and $H_c$ are constant with
increasing $n$.  In Fig. \ref{fig3}a), we plot three representative
data sets at $T=1, 10, 20$ K for convenient comparison.
For each temperature, a hysteresis loop is measured for a field
interval of $\pm 70$ mT.  Following the blue squares between $70$ and
$-5$ mT, the magnetization remains constant around $\mu_0 M=0.85$ T at
$T=1$ K.  At $-5$ mT the magnetization reversal process sets in by a
continuous decrease of magnetization by $0.19$ T followed by an abrupt
irreversible reversal step.  Then over a range of $3$ mT a plateau
forms which is followed by a smaller discontinuity in magnetization to
complete the magnetization reversal.  This second stage of reversal
vanishes at higher temperatures, resulting in a reversal which occurs
almost exclusively through a single irreversible change for both down-
and up-sweeps.  Proenca et al.\ observe a similar two-stage reversal
in an ensemble of Co NTs \citep{ProencaPRB:2013}, with the second
harder process also vanishing at high temperature.  They thus connect
this second stage of reversal to the exchange bias coupling. In
contrast to our observations on a single Py NT, the coherent rotation
measured in the Co/CoO arrays was extended over a much wider field
range.  Averaging over a distribution of different NTs in the array
provides a possible explanation for this difference.  In particular,
NTs in the array appear in a distribution of shapes and sizes.  In
addition, NTs in both studies are oxidized naturally in an
uncontrolled manner, allowing for a distribution of oxide thicknesses
within an array.  Since exchange bias crucially depends on film
characteristics, including graininess and thickness, the hysteresis
loops of the NT arrays are likely broadened by the distribution of
different NTs in the array.  For a detailed understanding,
measurements of single NTs are therefore critical.

At $T=20$ K (triangles) the hysteresis loop measured in the Py NT
shows perfectly symmetric behavior.  This symmetry, combined with the
observed reduction of $H_c$ and vanishing $H_{EB}$, indicates that the
NT has reached the Py/oxide system's blocking temperature $T_B$.
$\mu_0 H_c$ and $\mu_0 H_{EB}$ are plotted over the whole temperature
range in Fig.~\ref{fig3}b) allowing for the determination of $T_B
\approx 18$ K.  After $| H_{EB} |$ decreases with rising temperature
we find $\mu_0 H_{EB}=0 \pm 0.5$ mT above $12$ K, indicating a
blocking temperature $T_B$ in this regime.  A more precise
determination is possible, taking $\mu_0 H_c$ into account.  With
increasing temperature the coercivity shows a steady increase from
$14.5$ mT to $17$ mT until $16$ K. At $20$ K, the next investigated
temperature, $\mu_0 H_c$ drops below $14$ mT. This overall behavior is
in line with previous studies \citep{Kosub:2012} and allows us to
determine a blocking temperature $T_B=18 \pm 2$ K.

This very low blocking temperature deviates drastically from the bulk
values of the N\'{e}el temperatures $T_N$ of the possible native
oxides of Py, which are all well above $20$ K. This deviation suggests
that the oxide layer is very thin -- in the range of $3$ to $5$ nm --
and has a grainy and non-homogeneous structure \citep{Cortie:2012,
  NoguesNano:2005, NoguesEB:1999}. Indeed, similar blocking
temperatures around $T \approx 30$ K have been previously found for
naturally oxidized Py thin films \citep{Fulcomer:1972}.  The
decreasing extent of the continuous reversal region with increasing
temperature could be the result of the inhomogeneity and graininess of
the oxide shell.  Assuming grains of different dimensions, the
anti-ferromagnetic order could be gradually lost in different sections
of the shell with increasing temperature, thus leading to a gradual
temperature-dependent change in the reversal behavior.  Note that the
temperature dependence of the reversal process provides further
evidence that the exchange bias has a direct influence on the nature
of the magnetization reversal in the Py NT.

\begin{figure}[b]
  \centering
  \includegraphics[width=1\columnwidth]{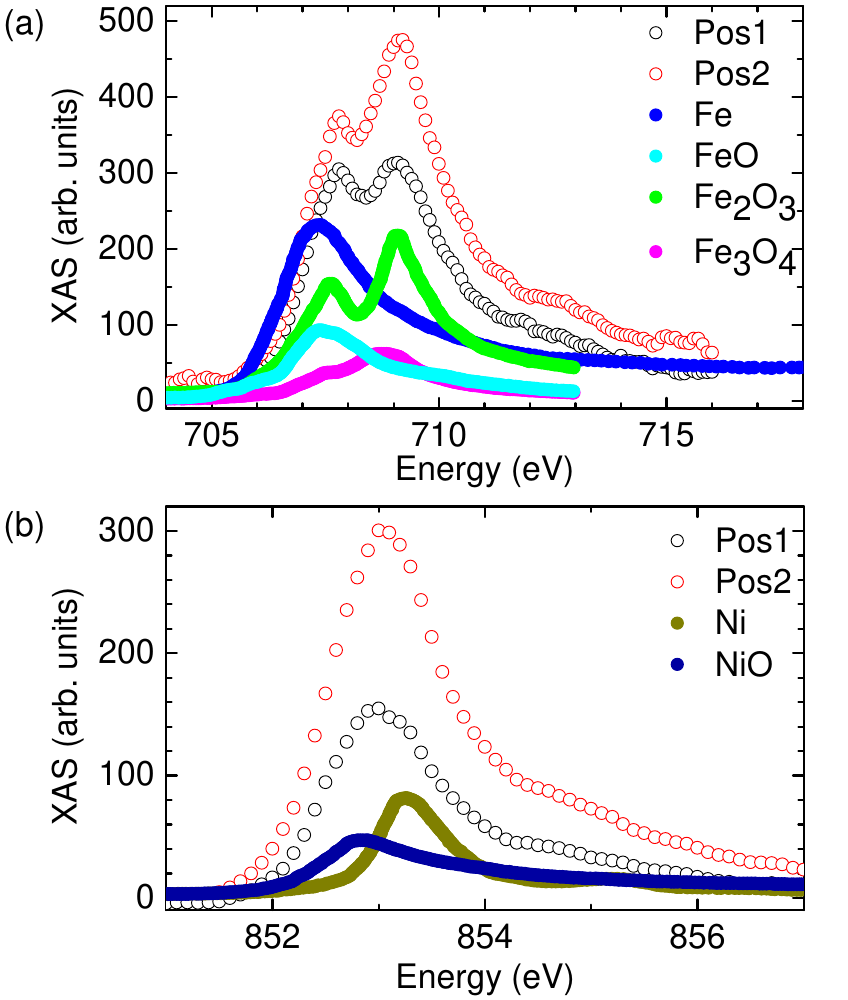}
\caption{
\label{fig4}
(a) Measured XAS of Fe L$_3$ edge and reference data for Fe and some
of its oxides taken from reference \citep{Regan:2001}. (b) Measured
XAS of Ni L$_3$ edge and reference data for Ni and its oxide NiO taken
from reference \citep{Regan:2001}.  }
\end{figure}

Given the measured exchange coupling, we investigate the nature of the
native oxide present on the Py NTs using spatially resolved X-ray
absorption spectroscopy (XAS) by means of X-ray photo-emission
electron microscopy (PEEM).  The experiments are performed using the
PEEM instrument at the surface/interface:microscopy beamline of the
Swiss Light Source at Paul Scherrer Institute
\citep{LeGuyader:2012}. NTs from the same growth wafer as the one used
in the magnetometry experiment are investigated. For the PEEM study
the NTs are transferred to a Si substrate with the aid of the
aforementioned micromanipulators and transferred into the
microscope. X-ray PEEM provides a spatial map of the local X-ray
absorption cross section. By recording such maps as a function of the
photon energy it is possible to record local X-ray absorption
spectra. Details are described, e.g., in
Refs. \citep{FraileRodriguez:2007, Vaz:2014}. The spatial resolution
of the instrument is between $50$ and $100$ nm and thus enables us to
perform XAS at various positions along the NT. We focus on XAS in the
vicinity of the L$_3$ edges of Fe and Ni to study possible oxidation
of the Py at different positions along the NT.  Two measurement
positions along a NT are chosen (at Pos 1: the NT's broken end and
another at Pos 2: close to the tip ). One additional section on the
substrate is chosen to allow for background subtraction.  The spectra
are recorded using circularly polarized X-rays. $\sigma^+$ polarized
light is used first, then after correcting for mechanical drift, the
polarization is changed to $\sigma^-$
\citep{FraileRodriguez:2010}. This procedure is repeated successively
for each position on the NT. Isotropic spectra are obtained by
averaging data according to $(\sigma^+ + \sigma^-)/2$.

Two representative spectra measured at the L$_3$ edge of Fe at Pos 1
and at Pos 2 are plotted in Fig. \ref{fig4}a).  The data resemble
typical spectra of oxidized Fe and thus suggest the presence of a Py
oxide layer on the Py NT.  They reveal Fe in different oxidation
states in the Py oxide shell in addition to metallic Fe in the Py
core, similar to what is observed in oxidized Fe surfaces,
cf. Ref. \citep{Vaz:2014}.  For comparison, reference data of pure Fe,
FeO, Fe$_2$O$_3$ and Fe$_3$O$_4$ taken from Regan et al.\
\citep{Regan:2001} are also plotted in Fig. \ref{fig4}a).  The
oxidation state of the present NT as seen at the Fe L$_3$ edge is
compatible with previous reports on comparable Py systems, which
identify FeO or $\alpha$-Fe$_2$O$_3$ in the native oxide of Py
\citep{Salou:2008,Fitzsimmons:2006}.  However, comparing the spectra
measured at Pos 1 and 2 further shows that the oxide layer is not
homogeneous in composition as indicated by the different ratio of the
peaks at $707.8$ and $709.2$ eV and by the varying signal amplitude,
which we assign to the differences in mechanical treatment when
picking up the NT.  In order to resolve the presence of NiO, we also
measure spectra at the Ni L$_3$ edge between $850$ and $858$ eV.  In
Fig.\ref{fig4}b) two representative curves are plotted for Pos 1 and 2
along with reference data for Ni and NiO from Regan et al.\
\citep{Regan:2001}.  The Py NT's data are compatible with a
superposition of spectra of NiO and metallic Ni, which suggests the
presence of a layer of NiO, in agreement with previous studies
\citep{Salou:2008,Fitzsimmons:2006}.

These results point to a layered composition of oxidized Py in the
following sequence: Py/($\alpha$-Fe$_2$O$_3$ or FeO)/NiO
\citep{Fitzsimmons:2006}.  All three oxides are antiferromagnetically
ordered below a certain ordering temperature (FeO: $198$ K,
$\alpha$-Fe$_2$O$_3$: $95$ K, NiO: $525$ K) in the bulk
\citep{Duo:2010}.  The overall thickness of the oxide layer is
estimated to be in the range of $2-5$ nm.  This estimate is based on
previous studies in literature \citep{Salou:2008,Fitzsimmons:2006} and
consistent with the fact that we detect a XAS signature from the
metallic Py in our sample with the typical probing depth of X-PEEM
being about $5$ nm.  In such thin layers magnetic order temperatures
are usually strongly reduced when compared to the bulk and thus might
explain why exchange bias in the present Py NTs is only observed at
temperatures below $20$ K.

In conclusion, in the absence of an exchange bias coupling with an
unintentional antiferromagnetic oxide shell, we find strong evidence
that the Py NT reverses its magnetization through the nucleation of a
vortex configuration at its end followed by an irreversible switching
process, as predicted by theory.  However, below $T_B \approx 18$ K
and before field training, we observe that the few-nanometer-thick
native oxide on the NT alters the process of magnetization reversal.
In particular, the non-equilibrium antiferromagnetic configuration of
the oxide appears to pin the magnetization of the NT and suppresses
the nucleation of the magnetic vortices at the NT ends for one of the
sweep directions.  Therefore, in order to control magnetization
reversal in Py NTs, one must control either the nature of the oxide
capping layer or work well above the determined blocking temperature
$T_B = 18$ K, where exchange bias is not effective.  These insights
come as a direct result of our hybrid magnetometer's ability to
measure both the behavior of the magnetic moments at the end of the NT
and the magnetization integrated throughout its volume.  Applying this
technique for the investigation of reversal processes in other types
of nanomagnets appears to be a promising path for future experiments.

\begin{acknowledgments}
  The authors thank Alan Farhan for technical assistance and
  acknowledge support from the Canton Aargau, the SNI, the SNF under
  Grant No. 200020-159893, the DFG via Projects No. GR 1640/5-2 in SPP
  1538 SpinCaT, KO 1303/13-1, KI 698/3-1, SFB TRR21 C2, and
  EU-FP6-COST Action MP1201.
\end{acknowledgments}

\end{document}